\documentclass[10pt,letterpaper,journal]{IEEEtran}
\usepackage[letterpaper,margin=1in]{geometry}
\usepackage{amsmath,amsfonts,amssymb}
\usepackage{algorithmic}
\usepackage{algorithm}
\usepackage{array}
\usepackage[caption=false,font=normalsize,labelfont=sf,textfont=sf]{subfig}
\usepackage{textcomp}
\usepackage{stfloats}
\usepackage{url}
\usepackage{verbatim}
\usepackage{graphicx}
\usepackage{cite}
\usepackage{xcolor}
\usepackage{newtxtext,newtxmath}
\usepackage{booktabs}
\usepackage[hidelinks]{hyperref}
\usepackage{bm}          
\DeclareUnicodeCharacter{2009}{\,}
\ifdefined\MARKEDVERSION
  \definecolor{revisionred}{RGB}{176,24,34}
\else
  \definecolor{revisionred}{RGB}{0,0,0}
\fi
\newenvironment{revision}{\begingroup\color{revisionred}}{\endgroup}
\newcommand{\rev}[1]{{\color{revisionred}#1}}
\newcommand{\mrev}[1]{\ifmmode\expandafter\mathcolor\else\expandafter\textcolor\fi{revisionred}{#1}}
\newcommand{\mkeep}[1]{\begingroup\color{black}#1\endgroup}
\hypersetup{pdftitle={A Unified Channel Model and Performance Evaluation for Underwater-to-Air Optical Links},pdfauthor={Xing Du and Mohamed-Slim Alouini},pdfsubject={Revised manuscript TCOM-TPS-26-1156}}
\begin{document}

\title{A Unified Channel Model and Performance Evaluation for Underwater-to-Air Optical Links}

\author{Xing Du and Mohamed-Slim Alouini,~\IEEEmembership{Fellow,~IEEE}
\thanks{This work was supported in part by the KAUST Office of Sponsored Research. \emph{(Corresponding author: Xing Du)}.}
\thanks{Xing Du was with the Computer, Electrical and Mathematical Sciences and Engineering Division, King Abdullah University of Science and Technology, Thuwal, Makkah 23955, Saudi Arabia. He is now with the School of Information Science and Engineering, Southeast University, Nanjing 210096, China (e-mail: duxing20@mails.ucas.ac.cn).}
\thanks{Mohamed-Slim Alouini is with the Computer, Electrical and Mathematical Sciences and Engineering Division, King Abdullah University of Science and Technology, Thuwal, Makkah 23955, Saudi Arabia (e-mail: slim.alouini@kaust.edu.sa).}%
}

\markboth{IEEE Transactions on Communications -- Revised Manuscript TCOM-TPS-26-1156}%
{Du \MakeLowercase{\textit{et al.}}: A Unified Channel Model and Performance Evaluation for Underwater-to-Air Optical Links}


\maketitle

\begin{abstract}
\rev{This paper develops a conditional statistical framework for nominally normal underwater-to-air optical links. It combines bulk attenuation, normalized log-normal atmospheric and Weibull underwater fluctuations, and entrance-pupil collection with surface-tilt-dependent Fresnel transmission. Conditioning transmission and beam displacement on a common tilt preserves their dependence. Product distributions with explicit support and Jacobians yield integrals valid for arbitrary positive Weibull shape. Average uncoded OOK bit error rate (BER) and outage are evaluated using consistent power, noise and SNR definitions. Independent component-wise Monte Carlo verifies the full conditional model at six reference configurations; targeted importance sampling checks selected BER values down to approximately $6\times10^{-9}$. A circular-pupil Gaussian integral reveals parameter-dependent errors in the equivalent-beam collection approximation, including a 36.9\% BER underestimate at one tested operating point. Conditional power and distance studies therefore use the circular-pupil reference and distinguish low-error short-air-segment conditions from collection-limited conditions. Source-anchored spectral and water-type contrasts, together with illumination and receiver checks, identify the dependence of these results on the prescribed operating conditions. The framework supplies reproducible error and outage calculations within a single-patch scalar model; physical operating ranges additionally depend on the surface, propagation and receiver assumptions.}
\end{abstract}

\begin{figure}[!t]
\centering\includegraphics[width=3.0in]{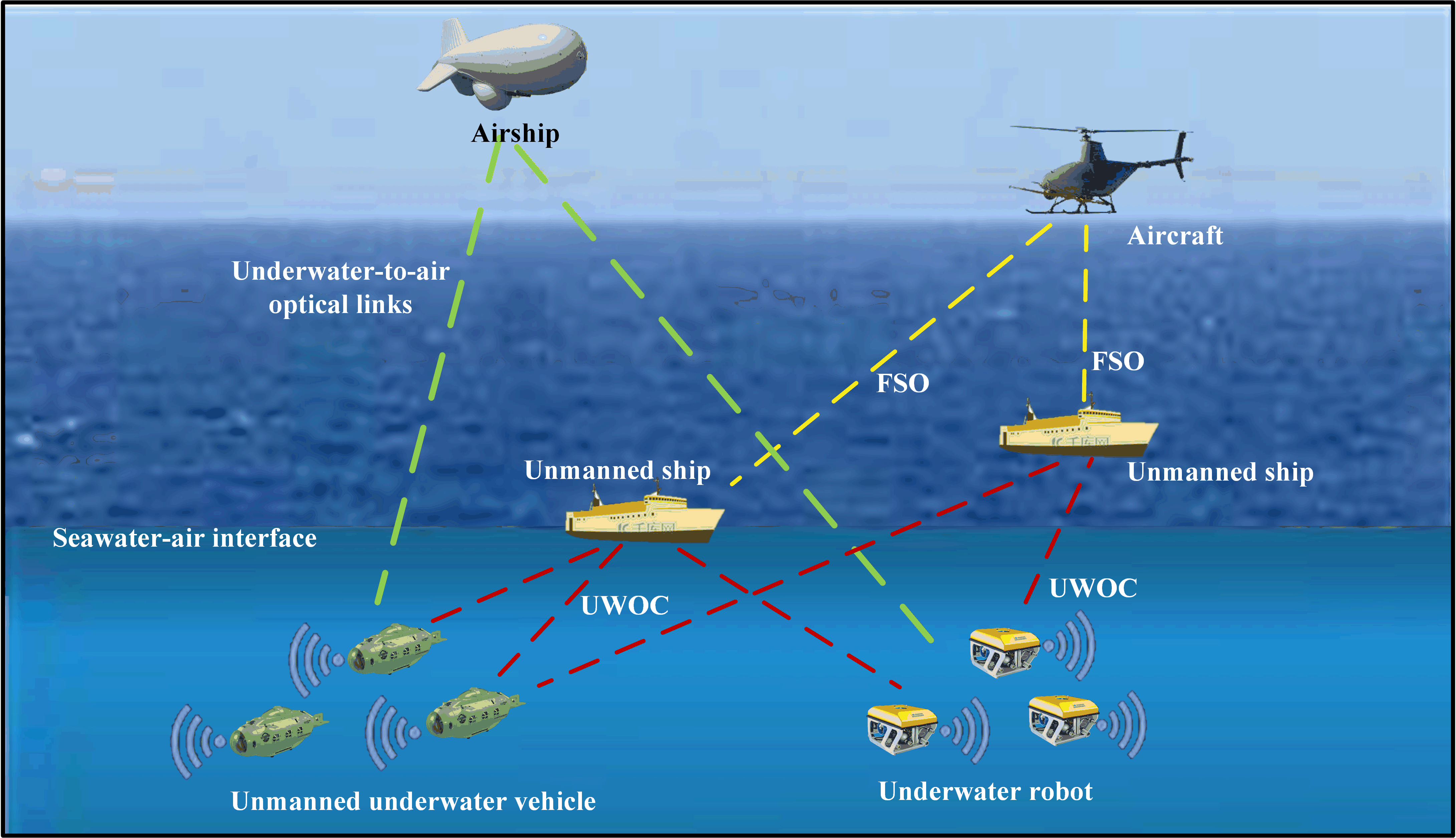}
\caption{\rev{Application schematic for underwater-to-air optical links. The derivation is restricted to a nominally normal single-patch uplink and a separately defined entrance pupil; the schematic is not a complete receiver layout.}}\label{fig1}
\end{figure}

\begin{revision}
\section{\mkeep{Introduction}}
\IEEEPARstart{D}{irect} optical transmission across the water--air interface can connect submerged sensors to surface or airborne receivers without an intermediate electrical relay. Field demonstrations show that wide-beam illumination and mobile receiver acquisition can support such links, but performance depends on water clarity, alignment and the actual optical receiver~\cite{ref_sun}. These demonstrations motivate a statistical link model; they do not validate an arbitrary combination of wavelength, divergence, aperture and noise parameters. Figure~\ref{fig1} illustrates the application context of the nominally normal uplink considered here.

Three distinctions are important when modeling this channel. First, bulk absorption and scattering extinction reduce the specified collected component, whereas normalized irradiance fluctuations change its instantaneous strength. Experimental fading studies show that the distribution depends on temperature, salinity and bubbles~\cite{ref_jamali}; using a Weibull family is therefore a conditional statistical choice, not a universal mapping from distance to turbulence. Second, an interface slope changes both Fresnel transmission and the outgoing direction. These effects share a physical cause and need not be independent. Third, an entrance pupil and a focal-plane detector are different collection planes. Beam divergence, pupil size, focal length and field of view cannot be interchanged in a power-loss formula.

The closest studies address different parts of this problem. Ata and Kiasaleh~\cite{ref10} characterize a direct seawater-to-air link using Beckmann pitching-angle statistics, log-normal underwater fluctuations and attenuation, and obtain gain statistics and outage performance. Wang \textit{et al.}~\cite{ref34} study random three-dimensional sea surfaces with Lambertian LED and simplified laser beam-spread descriptions. These are relevant direct-uplink comparisons, rather than studies that omit surface effects altogether. The short-range underwater thesis in~\cite{ref35} supplies historical context for optical-link design below the surface.

Underwater-only work by Boluda-Ruiz \textit{et al.}~\cite{ref_boluda} analyzes angular pointing, scattering-dependent geometric loss, Weibull fluctuations and intersymbol interference (ISI). Its geometric--pointing trade-off is established prior work, not a new effect claimed here. Likewise, the equivalent-beam collection approximation originates in existing pointing-error analysis~\cite{ref_farid}. Spatial propagation models provide another level of description: Gao \textit{et al.}~\cite{ref_gao} compare wave-spectrum-based sea surfaces using ray tracing and photon Monte Carlo. Such models can resolve surface structure and propagation effects that a single-patch scalar-gain model does not reproduce. The present analysis is complementary to, not a higher-fidelity replacement for, those calculations.

Recent direct-link work further narrows the relevant gap. Nennouche \textit{et al.}~\cite{ref_nennouche_e2e} combine photon transport, a JONSWAP surface, bubble scattering and UAV orientation with an LED/SiPM receiver. A subsequent preprint~\cite{ref_nennouche_beta}, updated in September 2026, fits a Beta-mixture pointing-loss distribution and includes total internal reflection and effective field-of-view interruption. Those loss mechanisms and the existence of interruption floors are therefore not claimed as new here. Unlike that fitted-mixture treatment, the present calculation integrates the specified slope/Rice law directly, retains continuous Fresnel transmission and independent unit-mean irradiance fluctuations, and tests the collection approximation at selected low-BER points. It does not provide their wind-to-UAV dynamics, fitted wind-regime calibration or photon-transport fidelity.

Relay systems must also be distinguished from direct optical transmission. The terrestrial--underwater amplify-and-forward (AF) system in~\cite{ref22} combines atmospheric and underwater optical hops with an intermediate receiver and retransmission. Its two-stage noise structure is not the single-detector noise model of a direct water--air crossing. The satellite--underwater acoustic relay study~\cite{ref_yang} similarly develops composite performance statistics, but its acoustic hop, optical satellite hop and AF relay do not constitute a matched optical-uplink baseline. Table~\ref{tab:related} compares the reported settings and analytical outputs. It does not infer an exhaustive absence of unlisted effects or claim a numerical reimplementation of different architectures.

\begin{table*}[!t]
\color{revisionred}\centering\footnotesize
\caption{\rev{Reported scope of selected related work and the present study. Comparisons describe the stated setting, not an accuracy ranking across different hardware or channel assumptions.}}\label{tab:related}
\renewcommand{\arraystretch}{1.05}\setlength{\tabcolsep}{4pt}
\begin{tabular}{@{}>{\raggedright\arraybackslash}p{.16\textwidth}>{\raggedright\arraybackslash}p{.19\textwidth}>{\raggedright\arraybackslash}p{.31\textwidth}>{\raggedright\arraybackslash}p{.28\textwidth}@{}}
\toprule
Study & Link setting & Reported model or evidence & Relation to the present scope\\\midrule
Ata--Kiasaleh~\cite{ref10} & Direct seawater--air & Beckmann pitching-angle and log-normal underwater statistics; PDF/CDF and outage. & Direct statistical comparator; present model conditions pupil collection and Fresnel transmission on a common tilt.\\
Wang \textit{et al.}~\cite{ref34} & Underwater uplink through random surface & Three-dimensional surface; LED Lambertian and laser beam-spread descriptions. & Prior random-surface uplink modeling; present outputs emphasize conditional gain and OOK error integrals.\\
Chancey~\cite{ref35} & Short-range underwater link & Underwater optical-link thesis. & Background, not an attributed water--air pitching-angle result.\\
Boluda-Ruiz \textit{et al.}~\cite{ref_boluda} & Underwater optical link & Scattering-aware geometric/angular loss, Weibull fading and BER/ISI analysis. & Established pointing/scattering trade-off; present scalar model does not inherit its temporal scattering treatment.\\
Gao \textit{et al.}~\cite{ref_gao} & Seawater--surface--air & Spatial wave spectra, ray tracing and photon Monte Carlo. & Propagation-fidelity reference; spatial surface/footprint effects are outside the single-patch model.\\
\'Alvarez-Roa \textit{et al.}~\cite{ref22} & Terrestrial--underwater AF relay & Atmospheric and underwater fading, geometric loss and relay/receiver noise; BER. & Different direction and noise architecture; not a direct-interface BER baseline.\\
Yang \textit{et al.}~\cite{ref_yang} & Acoustic--satellite optical AF relay & Acoustic and optical fading; outage, error and capacity-related analysis. & Cross-medium statistical context; no transplant of its SNR or distance results.\\
Sun \textit{et al.}~\cite{ref_sun} & Field water--air demonstration & Wide-beam illumination, prealignment and mobile signal acquisition. & Deployment evidence with different spectrum, modulation and receiver; not our test data.\\
Nennouche \textit{et al.}~\cite{ref_nennouche_e2e} & Direct LED--SiPM water--air & Photon transport, JONSWAP waves, bubble scattering and UAV orientation; end-to-end BER. & Higher-dimensional propagation and receiver model; different illumination/detector from the present scalar OOK reference.\\
Nennouche \textit{et al.}~\cite{ref_nennouche_beta} (preprint) & Vertical laser--UAV link & Wind/UAV statistics, fitted Beta-mixture pointing loss, TIR/FoV interruptions and outage. & Direct statistical comparator; no priority claim for interruption floors. Present integrals retain prescribed fluctuations without fitting the composite gain.\\
This work & Nominally normal direct optical uplink & Common-tilt Fresnel/Rice conditioning, unit-mean fluctuations, arbitrary-$K$ gain/BER integrals and circular-pupil comparison. & Reproducible scalar analysis with separated approximation, integration and sampling errors; source-anchored parameter contrasts.\\
\bottomrule
\end{tabular}
\end{table*}

The contribution is a conditional integration and assessment framework for a specified direct uplink, not a new fundamental fading distribution. It links the same local tilt to interface power transmission and pupil displacement; derives general-$K$ statistics and OOK performance with explicit supports; and separates the error of an inherited collection approximation from numerical verification of that approximation. A circular-pupil calculation and independent component-wise simulations quantify this distinction, including in selected low-BER tails. The resulting expressions are semi-analytical and are evaluated with explicitly specified numerical quadrature.

Source-anchored studies then distinguish an absorption-only spectral reference from published 532-nm water-type benchmarks, and an imposed broad envelope from a fixed-transmitter Gaussian reference. Receiver power conversion is tied to quantum efficiency, noise bandwidth and stated electrical inputs. These comparisons identify which conclusions follow from the scalar model and which depend on illumination, field of view or environmental calibration. They do not establish a universal preferred wavelength, a universal link range or a tested receiver implementation.

Section~II defines the model; Section~III develops gain, BER and outage statistics. Section~IV presents numerical verification and source-anchored contrasts, followed by conclusions in Section~V. The appendices provide transformations. The supplement records parameter provenance, additional sensitivities and numerical checks.
\end{revision}

\begin{revision}
\section{\mkeep{Channel Modeling}}\label{sec:model}
\subsection{Geometry, Reference Planes, and Assumptions}\label{sec:geometry}
We consider a quasi-static, nominally normal underwater-to-air uplink. The nominal underwater and air distances are $D_w$ and $D_a$, respectively. The transmitter illuminates one locally planar water--air patch; the receiver entrance pupil is centered on the nominal transmitted axis. The surface is represented by a random local slope, rather than by an independent exponential surface-height loss. Figure~\ref{fig1} illustrates possible applications, not a complete receiver optical layout. Oblique nominal incidence, surface shadowing, foam, and multiple interface encounters are outside the present single-patch model.

All transverse quantities used in the pointing-loss calculation refer to the \emph{receiver entrance-pupil plane}: $r$ is beam-centroid displacement, $\mkeep{r_a=D_G/2}$ is pupil radius, $\omega_z$ is the prescribed $1/e^2$ intensity radius in that plane, and $\sigma_s^2$ is the variance of each Cartesian component of residual displacement. It is not the variance of the radial displacement. The focal length $d_f$ is a separate optical quantity. An angular displacement produces a focal-plane shift approximately $d_f\tan\delta$, but that shift cannot be combined with entrance-pupil radii or beam widths. The present model assumes that the downstream detector collects all power passed by the pupil. A focal-length-dependent detector loss requires a separately specified focal-plane spot, detector size, and field stop; it is not implied by the pupil-loss model.

The channel power gain is written as
\begin{equation}
 H=LXYC,\qquad C=G(\eta)P,\qquad P=p(r),
 \label{eq:channel}
\end{equation}
where $X=h_{a,\mathrm{turb}}>0$ and $Y=h_{w,\mathrm{turb}}>0$ are normalized atmospheric and underwater irradiance fluctuations. $L$ contains only the deterministic nominal bulk attenuation. $G(\eta)$ accounts for interface transmission and the change in air-path length caused by refraction. The pupil collection factor $P$ and $G(\eta)$ depend on the \emph{same} surface tilt $\eta$ and are not assumed independent.

For fixed water properties, nominal geometry, and receiver configuration, $X$ and $Y$ are assumed mutually independent and independent of the local surface slope and residual receiver-plane displacement. These are conditional modeling assumptions, not consequences of the multiplication in \eqref{eq:channel}. They exclude correlated turbulence-induced beam wander and fluctuations in beam width or aperture averaging. A prescribed Gaussian beam envelope and a scalar irradiance fluctuation are used; conditions involving strong beam deformation, multiple scattering, or intersymbol interference require a different propagation model. Consequently, numerical agreement within this scalar model must be distinguished from experimental validation of these assumptions.

\begin{table}[!t]
\color{revisionred}
\caption{Model notation and units. Numerical inputs are specified in Tables~\ref{tab:parameters} and \ref{tab:source_inputs}.}\label{tab:notation}
\centering\footnotesize
\begin{tabular}{@{}p{0.23\columnwidth}p{0.70\columnwidth}@{}}
\hline
Symbol & Definition and unit \\ \hline
$D_w,D_a$ & Nominal water and air distances (m). \\
$\lambda,n_w,n_a$ & Vacuum wavelength (m); real refractive indices of water and air. \\
$c_w,\alpha$ & Water and air bulk extinction coefficients ($\mathrm{m}^{-1}$). \\
$\eta,\zeta,\delta$ & Local normal tilt, azimuth, and outgoing-ray deflection (rad). \\
$\sigma_w^2$ & Total mean-square slope; dimensionless, not a height variance. \\
$D_G,r_a$ & Entrance-pupil diameter and radius (m), $D_G=2r_a$. \\
$\omega_z,\omega_e$ & Pupil-plane Gaussian intensity radius and equivalent radius (m). \\
$v$ & Dimensionless pupil-to-beam ratio, $\sqrt\pi r_a/(\sqrt2\omega_z)$. \\
$\sigma_s,\sigma_g$ & Per-axis residual displacement standard deviation (m); Rayleigh tilt scale (rad). \\
$d_f$ & Focal length (m), not an entrance-pupil diameter. \\
$\vartheta_{\rm div}$ & Beam divergence half-angle (rad); not a length. \\
$A_0,P,C,L$ & Dimensionless power factors; for the equivalent-beam map, $0<P\leq A_0\leq1$, $0\leq C\leq A_0$. \\
$\sigma_l^2,K,\Lambda$ & Log-irradiance variance; Weibull shape and scale. \\
$P_t,R,\sigma_n^2$ & Mean launched OOK power (W), responsivity (A/W), decision-sample noise variance ($\mathrm{A}^2$). \\
$\gamma_0,\gamma,\mu_{\rm th}$ & Reference and instantaneous decision SNR; outage SNR threshold (all dimensionless). \\
\hline
\end{tabular}
\end{table}

\subsection{Bulk Attenuation and Interface Transmission}\label{sec:interface}
The nominal bulk attenuation is
\begin{equation} \begin{gathered}\mkeep{
 \mrev{L=h_{p,w,l}h_{p,a,l}},}\\\mkeep{
 h_{p,w,l}=e^{-\mrev{c_w(\lambda)}D_w},\quad
 h_{p,a,\mrev{l}}=e^{-\alpha(\lambda)D_a}.
 }\end{gathered}\label{eq:bulk}
\end{equation}
with $c_w(\lambda)=c_{\rm abs}(\lambda)+c_{\rm sca}(\lambda)$ and nonnegative extinction coefficients. This Beer--Lambert description concerns the ballistic or specified effective collected component~\cite{ref_boluda}. Including scattering in $c_w$ does not model the multiply scattered field, delayed photons, or the temporal impulse response. No general dominance of turbulence over scattering is assumed.

Let $m=n_w/n_a>1$, where the refractive indices correspond to the same optical frequency. For nominal normal incidence and a patch tilted by $\eta$, Snell's law gives
\begin{equation}
 \begin{gathered}
 \eta_t=\arcsin(m\sin\eta),\quad \delta(\eta)=\eta_t-\eta,\\
 \eta_c=\arcsin(1/m),
 \end{gathered}
 \label{eq:snell}
\end{equation}
for $0\leq\eta<\eta_c$. With real indices and a nonabsorbing interface, the power reflectances are~\cite{ref_fresnel}
\begin{equation}
 \begin{split}
 \mathcal R_s(\eta)&=\left(\frac{n_w\cos\eta-n_a\cos\eta_t}
 {n_w\cos\eta+n_a\cos\eta_t}\right)^2,\\
 \mathcal R_p(\eta)&=\left(\frac{n_a\cos\eta-n_w\cos\eta_t}
 {n_a\cos\eta+n_w\cos\eta_t}\right)^2.
 \end{split}\label{eq:fresnel_r}
\end{equation}
For unpolarized incident light, the interface power transmittance is
\begin{equation}
 T(\eta)=
 \begin{cases}
 1-[\mathcal R_s(\eta)+\mathcal R_p(\eta)]/2,&0\leq\eta<\eta_c,\\
 0,&\eta_c\leq\eta<\pi/2.
 \end{cases}\label{eq:fresnel_t}
\end{equation}
For prescribed polarization, the corresponding $s$- and $p$-power fractions replace the equal weights. Thus $0\leq T\leq1$, and the flat-interface limit is
\begin{equation}
 T_0=T(0)=\frac{4n_wn_a}{(n_w+n_a)^2}.
 \label{eq:flat}
\end{equation}
In particular, removing slope fluctuations does not remove the Fresnel loss. Surface-height variance and slope variance describe different surface properties; the latter determines the local incidence angle in this model.

For a receiver plane a nominal axial distance $D_a$ above the patch, the deflected central ray has air-path length $D_a/\cos\delta$. The factor in \eqref{eq:channel} is therefore
\begin{equation}
 G(\eta)=T(\eta)
 e^{-\alpha D_a[\sec\delta(\eta)-1]},\quad 0\leq\eta<\eta_c,
 \label{eq:ginterface}
\end{equation}
and $G=0$ for $\eta\geq\eta_c$. The exponential is only the \emph{additional bulk air attenuation}; the nominal factor is already in $L$. A small-angle reduction may set this exponential to one when $\alpha D_a[\sec\delta-1]\ll1$ over the retained operating range. Neither choice supplies a separate diffuse-scattering model. The Gaussian footprint approximation below additionally presumes that oblique footprint deformation is negligible over the relevant tilt range; large-angle predictions are formal extensions of the scalar model, not validated rough-sea predictions.

\subsection{Surface Statistics and Conditional Pointing Loss}\label{sec:pointing}
We use an isotropic Gaussian slope approximation to the Cox--Munk statistics~\cite{ref31}. If $s_x$ and $s_y$ are independent with variance $\sigma_w^2/2$, then $\tan\eta=(s_x^2+s_y^2)^{1/2}$ and the azimuth $\zeta$ is uniform on $[0,2\pi)$. The tilt density is
\begin{equation}\mkeep{ f_{\mrev{\eta}}(\eta)=\frac{2\tan\eta\mrev{\sec^2\eta}}{\sigma_w^2}
 e^{-\tan^2\eta/\sigma_w^2},\quad \mrev{0\leq\eta<\pi/2.}}\label{eq:tilt}
\end{equation}
Here $\sigma_w^2=\mathbb E[s_x^2+s_y^2]$ is a mean-square slope. The empirical wind relation $\mkeep{\sigma_w^2=0.003+0.00512U}$ is an optional empirical parameterization with $U$ in $\mathrm{m\,s^{-1}}$ under its source measurement convention; it is not a universal law for arbitrary wind-reference heights, swell, or breaking waves. The isotropic approximation omits directional slope statistics and surface shadowing. Alternatively, $\sigma_w^2$ can be specified from local slope measurements.

The inverse transformation
\begin{equation}\mkeep{ \eta=\arctan\sqrt{-\sigma_w^2\ln\xi},\qquad
 \mrev{\xi\sim\mathcal U(0,1),}}\label{eq:tilt_sample}
\end{equation}
is a sampling rule only. The auxiliary uniform variable $\xi$ is used only to generate samples from \eqref{eq:tilt}.

For a transmitted ray, let $d(\eta)=D_a\tan\delta(\eta)$ and let $\boldsymbol\epsilon\sim\mathcal N(\boldsymbol0,\sigma_s^2\boldsymbol I_2)$ denote independent residual displacement in the pupil plane. Then
\begin{equation}
 \boldsymbol r=d(\eta)(\cos\zeta,\sin\zeta)^{\mathsf T}
   +\boldsymbol\epsilon,\qquad r=\|\boldsymbol r\|.
 \label{eq:displacement}
\end{equation}
Conditioning on $\eta$ and integrating the azimuth of the residual displacement gives the Rice density
\begin{equation}
 f_{r\mid\eta}(r\mid\eta)=\frac{r}{\sigma_s^2}
 e^{-[r^2+d(\eta)^2]/(2\sigma_s^2)}
 I_0\!\left(\frac{rd(\eta)}{\sigma_s^2}\right),\quad r\geq0,
 \label{eq:rice}
\end{equation}
where $I_0$ is the modified Bessel function of the first kind. Every exponential and Bessel argument is dimensionless. The conditional mean displacement is fixed by the local refraction angle, while $\sigma_s$ describes the residual displacement per axis.

For the prescribed Gaussian pupil-plane envelope, we retain the equivalent-beam approximation~\cite{ref_farid}
\begin{equation}\mkeep{ \mrev{p(r)=A_0 e^{-2r^2/\omega_e^2}},\qquad
 A_0=\operatorname{erf}^2(v),\quad
 v=\frac{\sqrt\pi r_a}{\sqrt2\mrev{\omega_z}},}\label{eq:pupil}
\end{equation}
\begin{equation}\mkeep{ \omega_e^2=\omega_z^2
 \frac{\sqrt\pi\operatorname{erf}(v)}{2v e^{-v^2}}.}\label{eq:we}
\end{equation}
$A_0$ is the approximated on-axis collected fraction, not a turbulence normalization constant. This is an aperture approximation, not an exact diffraction calculation. An exact circular-aperture Gaussian integral is given in Appendix~\ref{app:pupil} as a reference for subsequent approximation checks. The value of $\omega_z$ must come from a consistent propagation or optical-design specification; it cannot be assigned independently of distance and divergence without explanation.

The inverse map and Jacobian are
\begin{equation}
 r(p)=\frac{\omega_e}{\sqrt2}\sqrt{\ln(A_0/p)},\qquad
 \left|\frac{dr}{dp}\right|=\frac{\omega_e^2}{4pr(p)}.
 \label{eq:p_jacobian}
\end{equation}
Consequently, for the equivalent-beam map,
\begin{equation}
 \begin{split}
 f_{P\mid\eta}(p\mid\eta)
 &=\frac{\omega_e^2}{4\sigma_s^2p}
 e^{-[r(p)^2+d(\eta)^2]/(2\sigma_s^2)}\\
 &\quad\times I_0\!\left(\frac{r(p)d(\eta)}{\sigma_s^2}\right),
 \quad 0<p<A_0.
 \end{split}\label{eq:p_conditional}
\end{equation}
The density is zero outside this support and integrates to one for each transmitted tilt when $\sigma_s>0$. At $\sigma_s=0$, the conditional law is a point mass at $p[d(\eta)]$, and expectations are evaluated with that limiting law instead of dividing by zero in \eqref{eq:p_conditional}.

For comparison with an independent first-order reduction, one may approximate $T$ by $T_0$, neglect the extra air-path attenuation, and linearize $\delta\simeq(m-1)\eta$. With a Rayleigh approximation for $\eta$, mean matching gives
\begin{equation}
 \sigma_g=\mathbb E[\eta]\sqrt{2/\pi},\qquad
 \sigma_p^2=\sigma_s^2+[D_a(m-1)]^2\sigma_g^2.
 \label{eq:rayleigh_scale}
\end{equation}
The direct small-slope approximation instead gives $\sigma_g^2\simeq\sigma_w^2/2$; these are alternative approximations, not exact interchangeable identities. Isotropic Rayleigh tilt with a uniform azimuth implies Gaussian Cartesian tilt components. Hence the reduced displacement is Rayleigh with per-axis variance $\sigma_p^2$, and
\begin{equation}
 f_P^{(0)}(p)=\frac{a}{A_0^a}p^{a-1},\quad
 a=\frac{\omega_e^2}{4\sigma_p^2},\quad 0<p<A_0.
 \label{eq:p_reduced}
\end{equation}
This reduced model is a comparison case only. Replacing $G(\eta)$ by a constant discards its dependence on pointing loss; the error of that replacement is not established by the algebraic reduction itself.

\subsection{Normalized Atmospheric and Underwater Fluctuations}\label{sec:turbulence}
Under the weak-atmospheric-fluctuation assumption~\cite{ref9},
\begin{equation} \begin{gathered}\mkeep{
 \mrev{\ln X\sim\mathcal N(-\sigma_l^2/2,\sigma_l^2),}}\\\mkeep{
 f_{\mrev{X}}(\mrev{x})=\frac{e^{-(\ln \mrev{x}+\sigma_l^2/2)^2/(2\sigma_l^2)}}
 {\mrev{x}\sqrt{2\pi\sigma_l^2}},\quad \mrev{x}>0.
 }\end{gathered}\label{eq:lognormal}
\end{equation}
Thus $\mathbb E[X]=1$ and $\mkeep{\mrev{\mathrm{SI}_a}=e^{\sigma_l^2}-1}$. Where the homogeneous plane-wave weak-fluctuation approximation applies, $\mkeep{\mrev{\mathrm{SI}_a}\simeq\sigma_R^2=1.23\mrev{C_n}^2\mrev{k_a}^{7/6}\mrev{D_a}^{11/6}}$, with $\mkeep{\mrev{k_a}=2\pi \mrev{n_a}/\lambda}$. A Gaussian-beam or aperture-averaged value should be obtained for the actual beam and pupil rather than identified automatically with this plane-wave estimate. $C_n^2$ is the refractive-index structure parameter, in $\mathrm{m}^{-2/3}$. The limit $\sigma_l=0$ is $X=1$ deterministically.

We retain the Weibull family for the underwater fluctuation~\cite{ref22,ref_boluda}, without imposing an inconsistent empirical shape--distance law:
\begin{equation} \begin{split}\mkeep{
 f_{\mrev{Y}}(\mrev{y})}&\mkeep{=\frac K\Lambda\left(\frac{\mrev{y}}\Lambda\right)^{K-1}
 e^{-(\mrev{y}/\Lambda)^K},\quad \mrev{y>0,}}\\\mkeep{
 \mrev{F_Y(y)}}&\mkeep{\mrev{=1-e^{-(y/\Lambda)^K}},\qquad K,\Lambda>0.
 }\end{split}\label{eq:weibull}
\end{equation}
Its moments give~\cite{ref_weibull}
\begin{equation}
 \begin{split}
 \mathbb E[Y^q]&=\Lambda^q\Gamma(1+q/K),\quad q>-K,\\
 \Lambda&=[\Gamma(1+1/K)]^{-1},\\
 \mathrm{SI}_w&=\frac{\Gamma(1+2/K)}{\Gamma(1+1/K)^2}-1.
 \end{split}\label{eq:weibull_moments}
\end{equation}
The stated choice of $\Lambda$ enforces $\mathbb E[Y]=1$, separating mean bulk attenuation from turbulence strength. A measured or otherwise justified scintillation index can determine $K$ through the last relation. The analysis uses the specified shape or a shape fitted to irradiance measurements; it does not assume a universal shape--distance relation. Values of $X$ or $Y$ above one represent normalized irradiance enhancements and do not conflict with the bounds on $T$ or $P$.
\end{revision}

\begin{revision}
\section{\mkeep{Performance Analysis of Cross-Medium Communication Systems}}\label{sec:performance}
\subsection{Scaled Atmospheric Gain and Conditional Product Law}\label{sec:product}
Let $H_1=LX$ with fixed $L>0$. For $x>0$,
\begin{equation}\mkeep{ f_{\mrev{H_1}}(\mrev{x})=\frac1{\mrev{L}} f_{\mrev{X}}(\mrev{x/L})
 =\frac{e^{-[\mrev{\ln(x/L)}+\sigma_l^2/2]^2/(2\sigma_l^2)}}
 {\mrev{x}\sqrt{2\pi\sigma_l^2}}.}\label{eq:h1}
\end{equation}
The density is expressed in the scaled-gain variable $x$. In the limit $\sigma_l=0$, $H_1=L$ and the corresponding probability measure is used.

Define $H_2=H_1G(\eta)P$. For a fixed transmitted tilt, $H_1$ and $P$ are independent under the assumptions of Section~\ref{sec:geometry}. The positive conditional density is
\begin{equation}
 \begin{split}
 f_{H_2\mid\eta}(u\mid\eta)
 &=\int_{u/[A_0G(\eta)]}^{\infty}
 \frac{f_{H_1}(x)}{xG(\eta)}\\
 &\qquad\times f_{P\mid\eta}\!\left(\frac{u}{xG(\eta)}\middle|\eta\right)dx,
 \quad u>0.
 \end{split}\label{eq:h2_conditional}
\end{equation}
Both the lower limit and the Jacobian follow from $0<P\leq A_0$. Averaging over the transmitted tilts gives
\begin{equation}
 f_{H_2}^{+}(u)=\int_0^{\eta_c}
 f_{H_2\mid\eta}(u\mid\eta)f_\eta(\eta)\,d\eta,
 \quad u>0.
 \label{eq:h2_positive}
\end{equation}
The superscript $+$ denotes the positive-density part, not a conditional density normalized to one. If the untruncated slope law in \eqref{eq:tilt} is used, total internal reflection gives
\begin{equation}
 p_0=\Pr(H_2=0)=\Pr(\eta\geq\eta_c)
 =e^{-\tan^2\eta_c/\sigma_w^2}.
 \label{eq:atom}
\end{equation}
Therefore $\int_0^\infty f_{H_2}^{+}(u)du=1-p_0$, and the full law includes $p_0\delta_0$. $\delta_0$ denotes unit probability mass at zero, not the deflection angle. The critical-angle tail is a mathematical consequence of the adopted untruncated slope model, not a measured blockage probability. In a study explicitly conditioned on transmission, the slope density on $[0,\eta_c)$ must instead be divided by $1-p_0$ and the conditional law has no atom; such conditional performance must not be reported as unconditional link availability.

For the independent reduced model in \eqref{eq:p_reduced}, define $L_0=LT_0$, $\widetilde H_1=L_0X$, and $\widetilde H_2=\widetilde H_1P^{(0)}$. Then the conventional product formula is recovered:
\begin{equation}\mkeep{ f_{\mrev{\widetilde H_2}}(\mrev{u})=\int_{\mrev{u/A_0}}^{\infty}
 \frac1{\mrev{x}} f_{\mrev{P}}^{\mrev{(0)}}(\mrev{u/x})f_{\mrev{\widetilde H_1}}(\mrev{x})\,d\mrev{x}.}\label{eq:h2_reduced_integral}
\end{equation}
Here $f_{\widetilde H_1}$ is \eqref{eq:h1} with $L$ replaced by $L_0$. In particular, a marginal $f_P$ cannot simply be multiplied by a random marginal interface transmittance in the coupled model.

A useful independent check of \eqref{eq:h2_reduced_integral} is available in the reduced case. Write $B=A_0L_0$, $b=u/B$, $s=\sigma_l$, and $E_a=e^{a(a+1)s^2/2}$. For $s>0$,
\begin{equation}
 f_{\widetilde H_2}(u)=\frac{a u^{a-1}}{B^a}E_a
 \overline\Phi\!\left(\frac{\ln b+(a+1/2)s^2}{s}\right),
 \label{eq:h2_reduced_pdf}
\end{equation}
\begin{equation}
 \begin{split}
 F_{\widetilde H_2}(u)
 &=\Phi\!\left(\frac{\ln b+s^2/2}{s}\right)\\
 &\quad+b^aE_a\overline\Phi\!\left(\frac{\ln b+(a+1/2)s^2}{s}\right),
 \end{split}\label{eq:h2_reduced_cdf}
\end{equation}
where $\Phi$ is the standard normal CDF and $\overline\Phi=1-\Phi$. These expressions are exact for the stated \emph{reduced} model, not for the tilt-coupled law. A stable implementation evaluates the normal survival function directly. Appendix~\ref{app:product} gives the underlying truncated log-normal moment.

\subsection{Composite Distribution for Arbitrary Weibull Shape}\label{sec:generalK}
The final gain is $H=H_2Y$. Since $Y>0$ almost surely and is independent of $H_2$, the mass at zero remains $p_0$. For $h>0$, conditioning on $H_2=u$ yields
\begin{equation}\mkeep{ f_{\mrev{H}}^{\mrev{+}}(h)=\int_{\mrev{0}}^\infty \frac1{\mrev{u}} \mrev{f_Y(h/u)}f_{\mrev{H_2}}^{\mrev{+}}(\mrev{u})\,d\mrev{u},}\label{eq:total_pdf}
\end{equation}
\begin{equation}
 F_H(h)=p_0+\int_0^\infty
 \left[1-e^{-(h/(\Lambda u))^K}\right]f_{H_2}^{+}(u)\,du.
 \label{eq:total_cdf}
\end{equation}
Equations~\eqref{eq:total_pdf}--\eqref{eq:total_cdf} hold for every $K>0$. No step assumes $K=2$, and no unexplained positive lower cutoff $h_M$ is introduced. For $h<0$, $F_H(h)=0$; at zero, $F_H(0)=p_0$. The nonnegative kernel implies monotonicity, and $F_H(h)\to1$ as $h\to\infty$. The reduced independent model uses $f_{\widetilde H_2}$ and $p_0=0$ in these expressions.

An equivalent CDF expression, useful without forming an intermediate product density, for $h>0$ is
\begin{equation}
 \begin{split}
 F_H(h)&=p_0+\int_0^{\eta_c}\!f_\eta(\eta)
 \int_0^\infty\! f_X(x)\int_0^{A_0}\! f_{P\mid\eta}(p\mid\eta)\\
 &\quad\times\left[1-e^{-\{h/[\Lambda LxG(\eta)p]\}^{K}}\right]
 dp\,dx\,d\eta.
 \end{split}\label{eq:direct_cdf}
\end{equation}
Its integrand is bounded between zero and one after conditioning. For positive real $q$, an additional moment identity is
\begin{equation}
 \mathbb E[H^q]=L^q e^{q(q-1)\sigma_l^2/2}
 \Lambda^q\Gamma(1+q/K)\mathbb E[C^q].
 \label{eq:total_moments}
\end{equation}
In particular, normalization of $X$ and $Y$ does not normalize the mean of the attenuated link gain to one.

\subsection{Numerical Quadrature and Endpoint Treatment}\label{sec:quadrature}
The preceding conditional integrals are exact probability relations within the specified model. Their finite-node evaluations are semi-analytical or quadrature-based results. The CDF can also be computed from
\begin{equation}\mkeep{ F_{\mrev{H}}(h)=\mrev{p_0+}\frac h2\int_{-1}^{1}
 f_{\mrev{H}}^{\mrev{+}}\!\left(\frac h2(1+\mrev{t})\right)d\mrev{t},\quad \mrev{h>0.}}\label{eq:cdf_transform}
\end{equation}
For an unweighted integral, first-kind Gauss--Chebyshev quadrature must include the weight conversion~\cite{ref_dlmf}:
\begin{equation} \begin{split}\mkeep{
 F_{\mrev{H}}(h)}&\mkeep{\simeq \mrev{p_0+}\frac{\pi h}{2\mrev{N_C}}
 \sum_{\mrev{j}=1}^{\mrev{N_C}}\mrev{\sqrt{1-t_j^2}}\,
 f_{\mrev{H}}^{\mrev{+}}\!\left(\frac h2(1+\mrev{t_j})\right),}\\\mkeep{
 \mrev{t_j}}&\mkeep{=\cos\left[\frac{(2\mrev{j}-1)\pi}{2\mrev{N_C}}\right].
 }\end{split}\label{eq:chebyshev}
\end{equation}
The factor $\sqrt{1-t_j^2}$ is essential because the native Chebyshev rule integrates with weight $(1-t^2)^{-1/2}$. Alternatively, Gauss--Legendre nodes $\ell_j$ and weights $v_j^{\rm L}$ directly approximate the unweighted integral:
\begin{equation}
 F_H(h)\simeq p_0+\frac h2\sum_{j=1}^{N_L}
 v_j^{\rm L}f_H^{+}\!\left(\frac h2(1+\ell_j)\right).
 \label{eq:legendre}
\end{equation}
Weights from different rules are not interchangeable. The bounded CDF kernel in \eqref{eq:total_cdf} or \eqref{eq:direct_cdf} is preferable when the positive density is sharply peaked or singular near zero. Log-domain integration and $-\operatorname{expm1}(-z)$ for $1-e^{-z}$ avoid avoidable underflow and cancellation. Node counts, integration tolerances, and convergence must be specified for each numerical result; they are not determined by these identities.

If integration over $u$ is restricted to $[u_-,u_+]$, the omitted contribution to \eqref{eq:total_cdf} is nonnegative and is at most
\begin{equation}
 \varepsilon_{\rm tail}=\int_0^{u_-}f_{H_2}^{+}(u)du
 +\int_{u_+}^{\infty}f_{H_2}^{+}(u)du.
 \label{eq:tail_bound}
\end{equation}
The atom $p_0$ must be retained separately. Truncating the low-gain tail without bounding its mass is not justified in outage or BER calculations.

\subsection{OOK Signal Convention and Average BER}\label{sec:ber}
We explicitly define $P_t$ as the \emph{average launched optical power} for equiprobable OOK, whose optical levels are $0$ and $2P_t$~\cite{ref_farid}. With a channel gain that is constant over the decision interval, the sampled photocurrent is
\begin{equation}
 Z=2RP_tH b+N,\qquad b\in\{0,1\},\quad
 N\sim\mathcal N(0,\sigma_n^2).
 \label{eq:ook_signal}
\end{equation}
$R$ is photodetector responsivity, not sensitivity; $\sigma_n^2$ is the variance of the decision-sample noise, not a noise spectral density. Background offsets are assumed removed. Noise is modeled as signal independent and equal for both symbols, and perfect knowledge of $H$ is assumed for the midpoint threshold $RP_tH$. Signal-dependent shot noise and channel-estimation errors are not included. If a data set specifies the on-state power $P_{\rm on}$, then $P_t=P_{\rm on}/2$ must be used.

For these conventions the conditional error probability and reference decision SNR are
\begin{equation}
 \begin{split}
 P_e(h)&=Q\!\left(\frac{RP_th}{\sigma_n}\right)
 =\frac12\operatorname{erfc}\!\left(\sqrt{\frac{\gamma_0}{2}}h\right),\\
 \gamma_0&=\left(\frac{RP_t}{\sigma_n}\right)^2,
 \qquad\gamma(H)=\gamma_0H^2.
 \end{split}\label{eq:ook_ber}
\end{equation}
This decision SNR is defined by the threshold separation and noise variance; it is not an unspecified optical SNR. The received mean decision SNR is
\begin{equation}
 \overline\gamma=\gamma_0\mathbb E[H^2],\qquad
 \gamma_{0,\rm dB}=10\log_{10}\gamma_0.
 \label{eq:snr_mean}
\end{equation}
An alternative, mean-square electrical-signal convention is $\gamma_{{\rm el},0}=2R^2P_t^2/\sigma_n^2=2\gamma_0$~\cite{ref_farid}. Under that convention the same physical conditional BER is $Q(\sqrt{\gamma_{{\rm el},0}/2}\,h)$, not $Q(\sqrt{\gamma_{{\rm el},0}}\,h)$. The numerical results use the decision-SNR convention in \eqref{eq:ook_ber}; conversion to an electrical-signal SNR must retain the corresponding factor of two.

Averaging the conditional BER, including possible zero gain, gives
\begin{equation}
 P_b=\frac{p_0}{2}+\int_0^\infty
 Q(\sqrt{\gamma_0}h)f_H^{+}(h)\,dh.
 \label{eq:ber_direct}
\end{equation}
Integration by parts with the full CDF, or equivalently averaging the integral representation of $Q$, yields
\begin{equation}
 P_b=\sqrt{\frac{\gamma_0}{2\pi}}
 \int_0^\infty e^{-\gamma_0h^2/2}F_H(h)\,dh.
 \label{eq:ber_cdf}
\end{equation}
The treatment of the atom is shown in Appendix~\ref{app:ber}. Setting $x=\sqrt{\gamma_0/2}\,h$ gives $h=\sqrt{2/\gamma_0}\,x$, and hence
\begin{equation}
 \begin{split}
 P_b&=\frac1{\sqrt\pi}\int_0^\infty
 e^{-x^2}F_H\!\left(\sqrt{\frac2{\gamma_0}}x\right)dx\\
 &=\frac1{2\sqrt\pi}\int_{-\infty}^{\infty}
 e^{-x^2}F_H\!\left(\sqrt{\frac2{\gamma_0}}|x|\right)dx.
 \end{split}\label{eq:ber_change}
\end{equation}
The CDF argument scales as $\gamma_0^{-1/2}$, not $\gamma_0^{-1}$. Both lines contain their required $1/\sqrt\pi$ normalization. With physicists' Hermite nodes $x_j^{\rm H}$ and weights $w_j^{\rm H}$ for the weight $e^{-x^2}$~\cite{ref_dlmf},
\begin{equation}\mkeep{ P_b\simeq\frac1{2\sqrt\pi}\sum_{\mrev{j}=1}^{\mrev{N_H}}\mrev{w_j^{\rm H}}
 F_{\mrev{H}}\!\left(\mrev{\sqrt{\frac2{\gamma_0}}|x_j^{\rm H}|}\right).}\label{eq:ber_hermite}
\end{equation}
These nodes and weights are different from the Chebyshev and Legendre sets. The transformed integrand can have a cusp at zero, so convergence must be checked against the half-line integral or direct conditional averaging. If $H=h_*>0$ deterministically, the exact result is $Q(\sqrt{\gamma_0}h_*)$; applying a finite-node rule across the resulting step CDF is not an exact AWGN check.

For a fixed gain distribution and the stated noise model, dominated convergence gives
\begin{equation}
 \lim_{\gamma_0\to\infty}P_b=\frac{p_0}{2}.
 \label{eq:ber_limit}
\end{equation}
Thus strictly positive multiplicative fading alone has no nonzero asymptotic BER floor. A nonzero floor in this particular extended law requires the explicitly modeled zero-gain probability. A plateau over a finite SNR sweep does not establish an asymptotic floor. Its interpretation requires a separate estimate of the zero-gain probability and an assessment of the slope model in the critical-angle tail.

\subsection{\mkeep{Outage Probability}}\label{sec:outage}
For a positive, dimensionless decision-SNR threshold $\mu_{\rm th}$,
\begin{equation}\mkeep{ P_{\rm out}=\Pr\{\mrev{\gamma_0H^2}\leq\mu_{\rm th}\}
 =F_{\mrev{H}}\!\left(\sqrt{\frac{\mu_{\rm th}}{\mrev{\gamma_0}}}\right).}\label{eq:outage}
\end{equation}
The CDF includes $p_0$ when the unconditional model is used. A threshold specified in dB is converted using $\mu_{\rm th}=10^{\mu_{{\rm th},\rm dB}/10}$. Receiver power sensitivity in watts is a different quantity and cannot be inserted into \eqref{eq:outage} without the defined photodetection and noise conversion. Equations~\eqref{eq:total_cdf}, \eqref{eq:direct_cdf}, and \eqref{eq:outage} provide a general-$K$ semi-analytical evaluation, not an elementary closed-form outage formula.
\end{revision}

\begin{revision}
\section{Numerical Verification and Reference Link Study}\label{sec:numerical}
\subsection{Configuration and Computational Procedure}\label{sec:configuration}
The numerical study uses the explicit reference configuration in Table~\ref{tab:parameters}. These are controlled model inputs rather than measurements of a deployed link. In particular, the water extinction coefficient is not assigned a named water class, and the slope and irradiance variances are specified directly rather than inferred from an unverified wind or distance fit. The reference wavelength is 525~nm. Sections~\ref{sec:source_inputs}--\ref{sec:illumination_receiver} separately introduce source-traced spectral, water-type, receiver and illumination contrasts; their assumptions are kept distinct from the verification benchmark.

To keep the beam radius and angular divergence consistent during an air-distance sweep, the prescribed paraxial Gaussian-intensity envelope is
\begin{equation}
 \omega_z(D_a)=\sqrt{\omega_I^2+(D_a\vartheta_{\rm div})^2},
 \label{eq:beam_envelope}
\end{equation}
where $\omega_I$ is the air-side interface-plane radius and $\vartheta_{\rm div}$ is a half-angle. Equation~\eqref{eq:beam_envelope} specifies an effective envelope, not a claim of diffraction-limited propagation. When $D_w$ varies, the interface illumination is held fixed: this isolates bulk underwater attenuation at a prescribed interface field. It does not describe varying water distance with an otherwise unchanged, fully specified transmitter optical train. The atmospheric log variance, underwater shape and slope variance are also held fixed within these controlled distance comparisons. A distance-independent $C_n^2$ is not simultaneously imposed: the plane-wave weak-fluctuation relation in Section~\ref{sec:turbulence} is a possible physical parameterization, whereas the controlled benchmark directly specifies $\sigma_l^2$.

\begin{table}[!t]
\color{revisionred}\centering\footnotesize
\caption{\rev{Reference inputs. All numerical assignments are declared modeling assumptions; the figure captions identify changes.}}\label{tab:parameters}
\begin{tabular}{@{}p{0.39\columnwidth}p{0.54\columnwidth}@{}}
\toprule
Quantity & Reference assignment \\\midrule
$D_w$; $D_a$ & 5~m; 2~m for distribution verification; air distance varied in the link study.\\
$\lambda$; $n_w,n_a$ & 525~nm; $1.34,1.0003$.\\
$c_w$; $\alpha$ & $0.15$; $2\times10^{-4}$~$\mathrm{m}^{-1}$.\\
$\sigma_w^2$; $\sigma_s$ & $0.003$; $5$~mm per Cartesian axis.\\
$\omega_I$; $\vartheta_{\rm div}$ & $40$~mm; $10$~mrad half-angle.\\
$D_G=2r_a$ & 40~mm entrance-pupil diameter.\\
$\sigma_l^2$ & $0.04$; $\mathrm{SI}_a=0.04081$.\\
$K$; $\Lambda$ & $1.5,2,5,10$ in verification; $K=5$ in the link study; $\Lambda=1/\Gamma(1+1/K)$.\\
$R$; $\sigma_n$ & $0.30$~A/W; $100$~nA decision-sample standard deviation.\\
$P_t$; $\mu_{\rm th}$ & 1 and 10~mW mean OOK power for the budget; $\mu_{\rm th}=10$ (10~dB).\\
Performance definition & Uncoded OOK; $\gamma_0=(RP_t/\sigma_n)^2$; ideal gain-aware threshold.\\
Numerical inputs & Component-wise MC: $10^6$ samples per verification case; node levels and seeds in Supplement~S1.\\
\bottomrule
\end{tabular}
\end{table}

Two pupil calculations are compared. The equivalent-beam map in \eqref{eq:pupil} is used to verify the transformed densities of Section~III. The circular-pupil integral in Appendix~A is then used as a reference for the same prescribed Gaussian envelope. It can also be evaluated as
\begin{equation}
 p_{\rm circ}(r)=F_{\chi'^2_2(4r^2/\omega_z^2)}
             \left(\frac{4r_a^2}{\omega_z^2}\right),
 \label{eq:pupil_ncx}
\end{equation}
because each Cartesian coordinate of a normalized Gaussian intensity footprint has variance $\omega_z^2/4$. Here $F_{\chi'^2_2(\nu)}$ denotes a noncentral chi-square CDF with two degrees of freedom and noncentrality $\nu$. This identity is checked against an independent polar integral, not against the equivalent-beam approximation. The maximum absolute discrepancy over 30 tested radius--offset pairs is $1.67\times10^{-16}$. ``Circular-pupil reference'' therefore means the exact collection integral for the prescribed envelope, not an exact wave-propagation or rough-surface solution.

For either pupil map $\psi(r)$, the CDF is computed without numerically differentiating a sampled distribution:
\begin{equation}
 \begin{split}
 F_H(h)=p_0+\int_0^{\eta_c}\!f_\eta(\eta)\int_0^\infty\!f_X(x)
 \int_0^\infty\!f_{r|\eta}(r|\eta)\\
 \times F_Y\!\left(\frac{h}{LxG(\eta)\psi(r)}\right)
 \,dr\,dx\,d\eta,\quad h>0.
 \end{split}\label{eq:radial_cdf}
\end{equation}
Thus using $p_{\rm circ}$ does not reuse the approximate pupil-density transformation as if it were exact. For the equivalent-beam case, \eqref{eq:radial_cdf} evaluates the same conditional product law as Section~III. BER is computed by conditioning on the intermediate gain $u=LxG(\eta)\psi(r)$ and averaging $\mathbb E_Y[Q(\sqrt{\gamma_0}uY)]$.

The slope transformation $q=\tan^2\eta/\sigma_w^2$ gives an exponential density. Composite Gauss--Legendre integration in $q$ and standardized Rice radius is combined with normal-weight Gauss--Hermite integration in $\log X$. Successive node levels use $(10,40,16)$, $(18,72,28)$, $(30,112,44)$ and $(44,160,64)$ nodes per slope panel, radial interval and normal integral, respectively. Panel endpoints, degeneracy handling and per-curve levels are supplied in Supplement~S1 and the executable configuration records. The ordinary checks retain $q\leq40$ and a nine-standard-deviation radial window; the omitted probability is bounded by $e^{-40}+e^{-81/2}=6.83\times10^{-18}$, independently of quadrature discretization error. We do not renormalize away this omitted mass. The atom at zero is evaluated separately.

\subsection{Independent Distribution and BER Verification}\label{sec:verification}
The Monte Carlo (MC) implementation generates the two Gaussian slope components and refracts the incident vector using the local surface normal. It independently generates the two residual displacement components, the log-normal atmospheric fluctuation and a unit-mean Weibull sample. It does not sample the final derived CDF. Six cases, each with $10^6$ realizations, test $K=1.5,2,5,10$ at $D_a=2$~m, $K=5$ at $D_a=50$~m, and a formal large-slope case with $\sigma_w^2=0.2$, $D_a=0.2$~m and $K=1.5$. The last case tests probability-mass accounting, not a validated rough-sea operating condition.

\begin{figure}[!t]
\centering\includegraphics[width=\columnwidth]{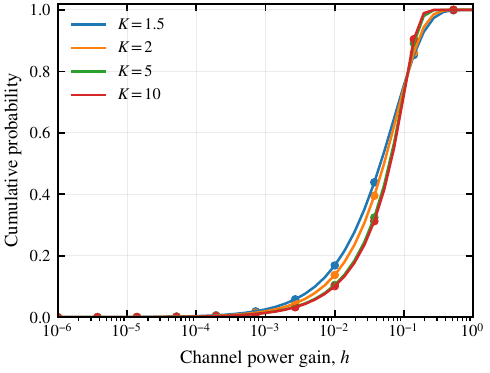}
\caption{\rev{CDF verification for the equivalent-beam model at $D_w=5$~m, $D_a=2$~m and four unit-mean Weibull shapes. Lines: conditional quadrature. Markers: independent component-wise MC, $10^6$ samples per curve. Other inputs are in Table~\ref{tab:parameters}.}}\label{fig2}
\end{figure}
\begin{figure}[!t]
\centering\includegraphics[width=\columnwidth]{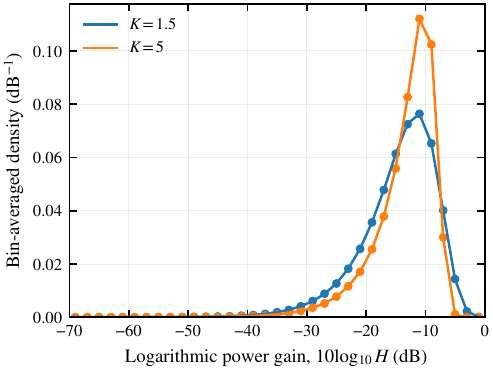}
\caption{\rev{Bin-averaged density of $10\log_{10}H$ for the equivalent-beam model at $D_w=5$~m and $D_a=2$~m. Lines are CDF differences divided by the dB-bin width, not point densities; markers are MC histogram frequencies divided by the same width.}}\label{fig3}
\end{figure}

Figure~\ref{fig2} compares the CDFs, and Fig.~\ref{fig3} checks bin probabilities in logarithmic gain coordinates. Across the six cases, the largest MC--quadrature CDF difference on the 43-point test grids is $6.38\times10^{-4}$. It is below the per-case 95\% Dvoretzky--Kiefer--Wolfowitz envelope $\sqrt{\log(40)/(2\times10^6)}=1.358\times10^{-3}$~\cite{ref_massart}. This is a grid comparison within a distribution-free envelope, not an empirical evaluation of the supremum over every gain value or a joint 95\% statement across all cases. The largest change between the two finest retained quadrature levels is $1.62\times10^{-7}$ in CDF. The largest tested positive-mass accounting error is $2.07\times10^{-14}$.

The base slope input gives $p_0=6.26\times10^{-183}$. In the formal large-slope test, $p_0=0.00184905$, compared with 1850 zero-gain events among $10^6$ MC samples. The latter verifies the atom separately from numerical underflow in very small positive gains. It does not establish an environmental blockage rate. Moment checks and all six case-level records are included in Supplement~S1.

Average conditional-noise BER is also checked against MC; the maximum standardized discrepancy on the ordinary verification grids is 3.07 estimated standard errors. At low BER, ordinary MC can miss the important tail events, even if its estimate looks smooth. We therefore use independent importance sampling for 18 selected points at $D_a=0.5$ and 1~m, with $K=5$, for both pupil maps. Each of the four distance--pupil groups uses $2^{20}$ samples split into four independent batches. The proposal mixes the nominal and broadened slope laws with nominal and lower-scale Weibull laws; explicit likelihood weights are retained. Supplement~S2 gives the distributions, seeds, weights and uncertainty estimates.

All 18 quadrature values differ from the importance-sampling estimates by at most 1.11 estimated standard errors; the largest relative standard error is 7.11\%. For example, with the circular pupil, $D_a=0.5$~m and $\gamma_{0,\rm dB}=52$, quadrature gives $5.92\times10^{-9}$ and importance sampling gives $(5.85\pm0.20)\times10^{-9}$, where the interval is an approximate 95\% normal interval. This is evidence for the numerical tail calculation under the specified model, not a measured bit-error count or post-FEC result. Independently evaluating the full CDF-based BER transform at four configurations agrees with direct conditional averaging to a maximum relative difference of $7.13\times10^{-10}$.

\subsection{Aperture Approximation and Link-Error Curves}\label{sec:aperture_comparison}
\begin{figure}[!t]
\centering\includegraphics[width=\columnwidth]{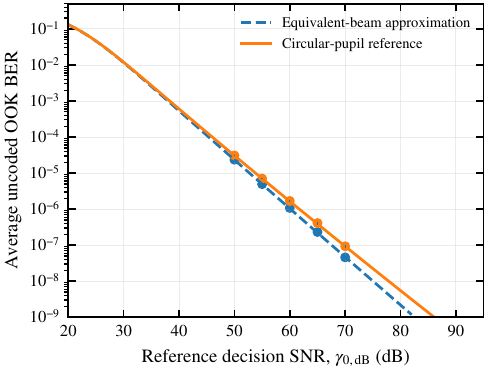}
\caption{\rev{Equivalent-beam approximation versus circular-pupil reference: $D_w=5$~m, $D_a=1$~m, $K=5$; all other inputs fixed at Table~\ref{tab:parameters}. Markers and approximate 95\% uncertainty bars at 50--70~dB are importance-sampling estimates. The horizontal axis is reference decision SNR, not received mean SNR.}}\label{fig4}
\end{figure}
Figure~\ref{fig4} shows why agreement between an approximate analytical model and its own simulator is not sufficient for design inference. At $D_a=1$~m and $\gamma_{0,\rm dB}=60$, the equivalent-beam map gives $P_b=1.0388\times10^{-6}$, whereas the circular-pupil reference gives $1.6471\times10^{-6}$. The former is 36.9\% lower, using the circular-pupil value as the denominator. The discrepancy grows in some lower-BER and larger-pupil settings (Supplement~S3). It is a collection-map approximation error, not a failure of the corrected product or BER identities. Consequently, the remaining design illustrations use $p_{\rm circ}$ rather than treating the equivalent-beam map as uniformly accurate.

\begin{figure}[!t]
\centering\includegraphics[width=\columnwidth]{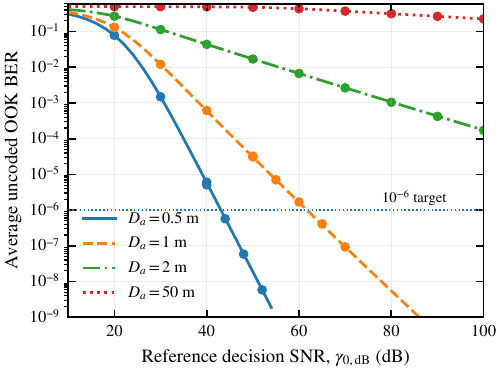}
\caption{\rev{Circular-pupil BER versus reference decision SNR at $D_w=5$~m and $K=5$, for four fixed air distances. Lines: quadrature. Markers: ordinary MC where resolved and importance sampling for selected low-BER points at 0.5 and 1~m. Values below $10^{-9}$ are not plotted. The $10^{-6}$ line is an illustrative uncoded target, not a universal standard.}}\label{fig5}
\end{figure}
Figure~\ref{fig5} exhibits both low-error short-air-segment conditions and high-error conditions for the same declared optical configuration. Increasing $D_a$ changes the refraction-induced centroid displacement and the prescribed beam radius, as well as bulk attenuation. This is not an isolated test of atmospheric extinction. Neither a universal underwater bottleneck nor a fixed water-distance cutoff follows from these curves. The SNR sweep is a mathematical reference sweep; interpreting it as an available launch-power range additionally requires a credible power/noise budget.

\begin{figure}[!t]
\centering\includegraphics[width=\columnwidth]{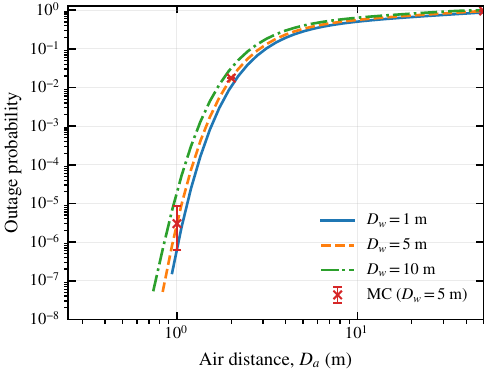}
\caption{\rev{Circular-pupil outage at mean launch power $P_t=1$~mW and decision-SNR threshold $\mu_{\rm th}=10$. Each curve varies $D_a$ only, with $D_w=1$, 5 or 10~m fixed; the interface envelope and fluctuation parameters are controlled as stated in Section~IV-A. Markers for $D_w=5$~m have binomial 95\% intervals. Zero observed outages are not plotted as zero true probability.}}\label{fig6}
\end{figure}
Figure~\ref{fig6} uses the same signal convention and pupil model. At $D_a=2$~m and $D_w=5$~m, quadrature gives outage 0.017553; an independent $10^6$-sample MC run gives 0.017580. At 50~m the corresponding values are 0.956080 and 0.956061. In contrast, observing zero outages among $10^6$ samples at 0.5~m provides an upper bound, not proof that the outage probability vanishes. Low-outage tails therefore require deterministic integration or targeted rare-event estimation rather than zero-count assertions.

\subsection{Conditional Budget, Sensitivities, and Design Scope}\label{sec:budget}
Table~\ref{tab:budget} converts the gain model to received power and uncoded BER using the assumed responsivity and decision-noise level. These assumptions do not specify receiver bandwidth, a data rate, background radiance or signal-dependent shot noise. The table is a conditional budget, not a hardware feasibility certification. The two launch powers correspond to $\gamma_{0,\rm dB}=69.54$ and 89.54, respectively.

\begin{table}[!t]
\color{revisionred}\centering\footnotesize
\caption{\rev{Conditional circular-pupil budget, $D_w=5$~m and $K=5$. $\overline P_r=P_t\mathbb E[H]$. All receiver and environmental assumptions are those of Table~\ref{tab:parameters}.}}\label{tab:budget}
\begin{tabular}{@{}rrrrr@{}}
\toprule
$D_a$ (m) & $P_t$ (mW) & $\overline P_r$ ($\mu$W) & BER & Outage\\\midrule
1 & 1 & 126.89 & $1.06\times10^{-7}$ & $3.02\times10^{-6}$\\
1 & 10 & 1268.91 & $<10^{-9}$\textsuperscript{a} & $1.04\times10^{-8}$\\
2 & 1 & 70.00 & $2.77\times10^{-3}$ & $1.76\times10^{-2}$\\
2 & 10 & 700.03 & $4.47\times10^{-4}$ & $2.78\times10^{-3}$\\
50 & 1 & 0.185 & 0.3754 & 0.9561\\
50 & 10 & 1.847 & 0.2686 & 0.6883\\\bottomrule
\end{tabular}
\par\smallskip\raggedright\scriptsize\textsuperscript{a}Computed below the displayed BER range; its precise magnitude is not independently validated by the selected rare-event tests.
\end{table}

For the 50-m air segment, \eqref{eq:beam_envelope} gives $\omega_z=0.502$~m, rather than a distance-independent 3-cm footprint. The on-axis circular collection fraction is only about 0.00317. The nominal air extinction is 0.0434~dB, the nominal water loss is 3.257~dB, and the flat-interface loss is 0.0925~dB. Surface-induced displacement further reduces the average collection. These numbers explain why increasing power from 1 to 10~mW does not bring this small, untracked pupil close to the target. They do not show that 50-m air links are impossible with different illumination, tracking, apertures or receiver noise. Conversely, the successful short-air-segment example does not certify a particular platform or data rate.

Supplement~S3 varies mean-square slope, residual displacement, underwater Weibull shape, atmospheric log variance, pupil diameter and the water extinction coefficient while holding the other inputs fixed. These are controlled parameter axes; source-traced spectral and named-water contrasts are reported separately in Section~\ref{sec:spectral_water}. Normalizing $X$ and $Y$ to unit mean prevents a turbulence-strength change from silently changing mean bulk loss. A separate ablation with $X=1$, $Y=1$, or both removes irradiance fluctuations while retaining surface motion, pointing and water extinction. It measures their conditional impact on this scalar link; it does not compare multiply scattered collection or intersymbol interference against turbulence.

A same-marginal permutation test also separates unconditional independence from the physical common-tilt model. It keeps the distributions of $G$ and $P$ but permutes one factor across samples. The change depends on the slope regime and the tested operating point; small ordinary-MC estimates with large relative uncertainty are explicitly marked unresolved. The larger-slope comparisons are formal stress tests. A small dependence effect at one reference setting is not a general justification for replacing the joint law by independent marginals.

The required-power curves in Supplement~S3 invert the circular-pupil BER for the $10^{-6}$ target. For $D_w=5$~m, the assumed receiver requires approximately 0.407~mW at $D_a=1$~m. An illustrative 10-mW cap is used only to distinguish the displayed conditional operating region from formal extrapolation; it is not a safety or regulatory power limit. Requirements above this cap are not proposed equipment specifications. The following sections distinguish published optical-property benchmarks from the remaining requirements for deployment-specific calibration.

\subsection{Source-Traced Optical and Receiver Inputs}\label{sec:source_inputs}
Three input categories are kept separate: the controlled verification benchmark in Table~\ref{tab:parameters}, published optical-property values, and declared receiver/illumination assumptions. The contrasts test parameter dependencies without assigning the benchmark to a measured device. Water index follows Quan--Fry~\cite{ref_quan} within its temperature, salinity and wavelength domain, using $n_a=1$ as an optical approximation. The air extinction, local slope and scalar fluctuation statistics remain fixed controlled inputs; no atmospheric spectral calibration or wind-to-turbulence calibration is inferred.

The PIN-like receiver uses the responsivity and noise relations~\cite{ref_hama}
\begin{align}
 R(\lambda)&=\eta_q e\lambda/(h_{\rm P}c_0),\label{eq:responsivity_spectral}\\
 \sigma_0^2&=B_e\!\left[i_{\rm amp}^2+\frac{4k_{\rm B}T_e}{R_L}
       +2e\{I_d+R(\lambda)P_{\rm bg}\}\right],\label{eq:receiver_noise}\\
 \gamma_{0,\mathrm{eff}}&=\left[R(\lambda)\eta_{\rm opt}P_t/\sigma_0\right]^2.
 \label{eq:snr_eff}
\end{align}
Here $h_{\rm P}$ is Planck's constant, $c_0$ is vacuum light speed and $e$ is the elementary charge. $\eta_{\rm opt}$ is throughput after the modeled pupil, so it is not already in $H$. The gain definitions in Sections~II--III are unchanged; only the explicitly defined receiver power conversion changes. $R_L$ is an input-equivalent resistive noise element, not an asserted diode shunt resistance. The white input-current density $i_{\rm amp}$ summarizes electronics noise. $B_e$ is equivalent noise bandwidth, not a stated bit rate. Equal OOK probabilities and the gain-aware midpoint detector remain assumed.

\begin{table}[!t]
\color{revisionred}\centering\footnotesize
\caption{\rev{Inputs for source-traced contrasts, distinct from the verification benchmark. Published values and design assumptions are identified explicitly.}}\label{tab:source_inputs}
\begin{tabular}{@{}p{.39\columnwidth}p{.54\columnwidth}@{}}
\toprule
Item & Assignment and provenance\\\midrule
Spectral water reference & Pope--Fry pure-water absorption at 450--550 nm~\cite{ref_pope}; $T=22^{\circ}$C, $S=0$. No scattering spectrum added.\\
Named-water contrast & 532-nm clear-ocean/coastal $(a,b,c)$ from~\cite{ref_boluda}, Table 1; index uses $T=20^{\circ}$C, $S=35$\,\textperthousand.\\
Index & Quan--Fry equation~\cite{ref_quan}; $n_a=1$ approximation.\\
$\eta_q$; $\eta_{\rm opt}$ & $0.70$; $0.80$: assumed efficiencies, not device specifications.\\
$T_e$; $R_L$; $i_{\rm amp}$ & 300 K; $50\,\Omega$; $26$ pA/$\sqrt{\mathrm{Hz}}$: declared electrical inputs.\\
$I_d$; $P_{\rm bg}$; $B_e$ & 1 nA; 0; 10 MHz: dark-reference inputs, not a daylight receiver.\\
525-nm derived values & $R=0.2964$ A/W; $\sigma_0=100.37$ nA.\\
Other channel inputs & Table~\ref{tab:parameters}; $K=5$, circular pupil, $D_a=1$ m unless stated.\\
\bottomrule\end{tabular}
\end{table}

Table~\ref{tab:source_inputs} gives a reproducible origin for approximately 100-nA sample noise, under the declared receiver assumptions. At 525 nm, 10 MHz, $D_w=5$ m and $c_w=0.15$~m$^{-1}$, the $10^{-6}$ BER target requires 0.544 mW with the new throughput and responsivity. This receiver configuration is distinct from the controlled benchmark in Table~\ref{tab:parameters}. At otherwise fixed inputs, required power scales as $\sqrt{B_e}/\eta_{\rm opt}$ (Supplement~S7). Finite detector area, capacitance, bandwidth and field of view must still be jointly realized in hardware; a PIN-noise calculation does not establish that optical implementation.

Signal-dependent shot noise is checked at six selected $10^{-6}$ operating points using a paired importance estimator and the same midpoint threshold. The largest estimated relative BER increment is less than $0.008\%$, including a pointwise approximate uncertainty allowance. Supplement~S7 gives the unequal-variance conditional error law. This finite check supports the equal-noise approximation at those dark-reference points; it does not cover arbitrary powers, daylight backgrounds or an optimized unequal-noise detector.

\subsection{Spectral Reference and Published Water-Type Contrasts}\label{sec:spectral_water}
The spectral calculation uses pure-water absorption $a_w=0.00922$, $0.0114$, $0.0204$, $0.0417$ and $0.0565$~m$^{-1}$ at 450, 475, 500, 525 and 550 nm, respectively, from~\cite{ref_pope}, Table 3. Setting $c_w=a_w$ deliberately defines an absorption-only lower-loss reference; it is not a measured total-extinction spectrum for pure or coastal seawater. Index, refraction and responsivity change with wavelength, whereas the imposed interface envelope, fluctuation laws, $\eta_q$ and $\eta_{\rm opt}$ are held fixed. This separates spectral-property effects from an unmodeled spectral change of source optics or a real detector's efficiency curve.

\begin{figure}[!t]\centering\includegraphics[width=\columnwidth]{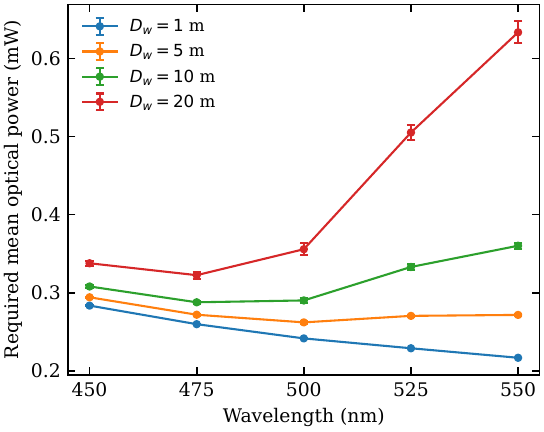}
\caption{\rev{Required mean OOK power for $10^{-6}$ BER in the absorption-only spectral reference; $D_a=1$ m, circular pupil and Table~\ref{tab:source_inputs}. Bars perturb absorption by its published one-standard-deviation value with all other inputs fixed; they are not total confidence intervals. Lines connect five evaluated wavelengths, not a continuous wavelength optimization.}}\label{fig7}\end{figure}

Figure~\ref{fig7} shows that 525 nm is not a universal optimum. At $D_w=20$ m, required mean powers at 475, 525 and 550 nm are approximately 0.322, 0.505 and 0.634 mW, respectively. Among the five nominal grid values, the minimum changes with distance because absorption and photoelectric conversion compete. Near-equal grid values should not be ranked as robust optima: at 10 m, the 475- and 500-nm absorption-perturbation ranges overlap. For a change $\Delta a_w$ alone, $P_{t,\rm req}$ is multiplied exactly by $\exp(D_w\Delta a_w)$ under the fixed-gain-factor assumptions; the plotted bars propagate only this one source of uncertainty.

A separate 532-nm comparison uses the clear-ocean benchmark $(a,b,c)=(0.114,0.037,0.151)$ and coastal benchmark $(0.179,0.219,0.398)$~m$^{-1}$ from~\cite{ref_boluda}. The nominal water classes are those of the cited benchmark, not measurements at our site and not a 525-nm relabeling. With all other factors fixed,
\begin{equation}
 \frac{P_{t,\rm req}(c_2)}{P_{t,\rm req}(c_1)}
       =\exp[(c_2-c_1)D_w].\label{eq:water_penalty}
\end{equation}
This familiar attenuation identity is used as a consistency check, not claimed as a new propagation result.

\begin{figure}[!t]\centering\includegraphics[width=\columnwidth]{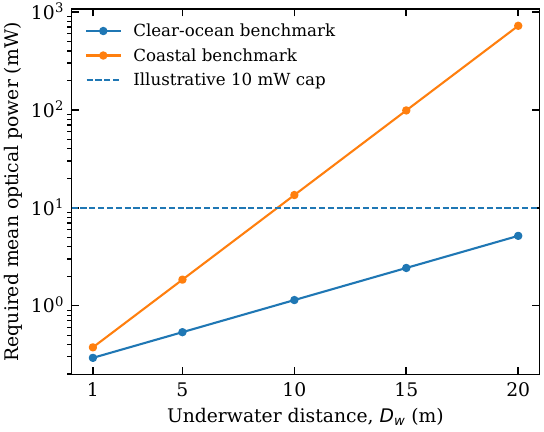}
\caption{\rev{Published 532-nm water-coefficient contrast at $D_a=1$ m and $10^{-6}$ target BER. Only bulk water extinction changes within each controlled distance sweep; the interface envelope and fluctuations are fixed. Large powers and scattering optical depths are formal ballistic/effective-component extrapolations, not deployable-link forecasts. The 10-mW cap is illustrative, not a safety limit.}}\label{fig8}\end{figure}

At $D_w=10$ m, Fig.~\ref{fig8} gives 1.140 and 13.483 mW, a coastal/clear power penalty of about 10.73 dB. The corresponding scattering optical depths $bD_w$ are 0.37 and 2.19. The latter warns against interpreting the scalar result as the full collected scattered field: multiple scattering, footprint change and ISI require transport/temporal modeling~\cite{ref_boluda,ref_nennouche_e2e}. Thus the benchmark quantifies the specified component's sensitivity, rather than asserting that turbulence dominates coastal-water scattering.

All 30 wavelength--distance and water-type operating points were recomputed. Two finest retained quadrature levels change the evaluated target BER by less than $7.5\times10^{-10}$ relatively for this finite grid. Six independent importance checks, each with $2^{20}$ samples in four batches, agree with their $10^{-6}$ targets within 1.43 estimated standard errors; the largest relative standard error is 4.19\%. These tests concern computation under the stated inputs, not uncertainty in the empirical water coefficients or the real channel.

\subsection{Illumination and Receiver Acceptance Requirements}\label{sec:illumination_receiver}
The 40-mm interface radius and 10-mrad half-divergence in the verification benchmark must not be described as a diffraction-limited Gaussian laser. Were that plane treated as a waist, the beam-parameter product would imply $M^2=\pi n_a\omega_I\vartheta_{\rm div}/\lambda\simeq2.39\times10^3$. This is a consistency diagnosis of the imposed envelope, not a measured beam-quality factor. Wide-beam field demonstrations~\cite{ref_sun} establish the relevance of broad illumination, but do not supply these exact input values.

For comparison, a fixed underwater transmitter waist $w_0$, a normal mean interface and paraxial Gaussian propagation give~\cite{ref_gaussian}
\begin{equation}
 \omega_z^2=w_0^2+
 \left[\frac{M^2\lambda}{\pi w_0}
       \left(\frac{D_w}{n_w}+\frac{D_a}{n_a}\right)\right]^2.
 \label{eq:two_medium_gauss}
\end{equation}
The reduced-distance derivation is given in Supplement~S6. We evaluate $M^2=1$ and $w_0=20$, 40 and 80 mm at five air distances, together with the broad-envelope benchmark. Random tilt shifts the prescribed footprint; fluctuating beam deformation and multiple-scattering broadening remain excluded. These are illumination contrasts, not optimized tracking or platform designs.

A narrower beam is not necessarily better for an untracked pupil. At 10-mW mean power and $D_a=2$ m, the 20- and 80-mm fixed-waist references give BERs of approximately $5.32\times10^{-2}$ and $6.55\times10^{-8}$, respectively (Fig.~S10). At 50 m, all four tested illumination choices remain above 0.28 under the new receiver assumptions. These comparisons separate aperture dilution from pointing tolerance; they do not establish a general benefit of scattering or a universal range ceiling.

The assumption that the detector accepts all power admitted by the pupil also needs an angular check. For an ideal added hard field stop, a chief-ray miss probability $p_{\rm miss}$ contributes at least $p_{\rm miss}/2$ to average OOK error. Under the adopted slope law, the required half-angle obeys
\begin{align}
 \eta_*&=\arctan\sqrt{-\sigma_w^2\ln p_{\rm miss}},\nonumber\\
 \psi_{\rm FOV}&\geq\arcsin(m\sin\eta_*)-\eta_* .\label{eq:fov_requirement}
\end{align}
For $p_{\rm miss}=2\times10^{-6}$, the 525-nm reference needs approximately $3.92^{\circ}$ half-FoV, or at least a 3.42-mm chief-ray radius at a 50-mm focal length. This is only a necessary acceptance condition: equality already consumes the entire $10^{-6}$ error budget, and focused-spot size, residual angles and design margin require further allowance. No finite-FoV detector has been silently inserted into the earlier performance curves. Existing TIR/FoV studies~\cite{ref_nennouche_beta} reinforce why acceptance cannot be omitted from a deployment claim. The present result identifies a receiver requirement to be checked, not an achieved hardware capability.

\end{revision}

\begin{revision}
\section{\mkeep{Conclusions}}\label{sec:conclusion}
A common-slope formulation couples water--air Fresnel transmission and refraction-induced pupil displacement while keeping nominal bulk attenuation and normalized irradiance fluctuations distinct. The product transformations respect the pupil-gain support, the Weibull integration is valid for arbitrary $K>0$, and the OOK definitions lead to a consistent inverse-square-root SNR transformation and numerical quadrature.

The full conditional model is checked against independent component-wise Monte Carlo at six reference configurations, and selected low-BER points are additionally checked by importance sampling. These checks establish numerical consistency within the adopted model. They also reveal a separate limitation: the equivalent-beam pupil approximation can be appreciably optimistic near a target BER, even when its own analytical and simulated distributions agree. Circular-pupil reference calculations are therefore used for the conditional operating-region analysis.

The source-traced studies qualify the engineering interpretation. Pure-water absorption and idealized photoelectric conversion do not select one preferred wavelength at every distance. Published 532-nm clear/coastal coefficients produce markedly different conditional power budgets, while the coastal scattering optical depth limits a scalar interpretation. A fixed-transmitter Gaussian reference shows why narrowing an untracked beam does not automatically improve BER. Explicit noise bandwidth and field-of-view requirements replace undefined receiver sensitivity and focal-length benefits.

The base slope law gives negligible zero-gain mass, so its high finite-SNR BER is not evidence of an asymptotic floor. TIR and finite-FoV interruption mechanisms have already been studied in related work; the present contribution is the conditional, general-$K$ performance calculation and its separation of collection approximation from numerical validation. The 50-m air case remains unsupported by the tested untracked configurations, not ruled out for all designs. Measured surface/irradiance statistics, complete scattering spectra, spatial propagation and a jointly realizable finite-detector receiver are required for deployment-specific predictions.
\end{revision}

\appendices
\begin{revision}
\section{Pupil Collection and Conditional Displacement}\label{app:pupil}
For the normalized Gaussian intensity $2e^{-2\|\boldsymbol\rho-\boldsymbol r\|^2/\omega_z^2}/(\pi\omega_z^2)$ in a circular pupil, the exact collected fraction of that prescribed profile is
\begin{equation}
 P_{\rm circ}(r)=\frac4{\omega_z^2}\int_0^{r_a}
 \rho e^{-2(\rho^2+r^2)/\omega_z^2}
 I_0\!\left(\frac{4\rho r}{\omega_z^2}\right)d\rho.
 \label{eq:exact_pupil}
\end{equation}
In particular, $P_{\rm circ}(0)=1-e^{-2r_a^2/\omega_z^2}$. This differs from the approximate $A_0$ in \eqref{eq:pupil}; agreement must be checked for the adopted beam-to-pupil ratio. This reference integral does not add multiple scattering or a receiver-lens diffraction model.

For a two-dimensional Gaussian displacement with fixed mean vector of magnitude $d$, polar integration uses
\begin{equation}
 \frac1{2\pi}\int_0^{2\pi}e^{z\cos\varphi}\,d\varphi=I_0(z),
 \label{eq:rice_identity}
\end{equation}
which gives \eqref{eq:rice}. Substituting the monotone pupil map and the Jacobian in \eqref{eq:p_jacobian} then proves \eqref{eq:p_conditional}. No independent angular marginalization of $T(\eta)$ is involved.

\section{Product Support, Reduced Check, and General-\texorpdfstring{$K$}{K} Integration}\label{app:product}
If $V=X_1P$ with $X_1>0$ independent of $P\in(0,A_0]$, then, at fixed $X_1=x$, the transformation is $p=v/x$ and $dp=dv/x$. The admissibility condition $v/x\leq A_0$ implies $x\geq v/A_0$. This proves \eqref{eq:h2_reduced_integral}; replacing $x$ by $xG(\eta)$ while conditioning on $\eta$ proves \eqref{eq:h2_conditional}.

For $Z=\ln X\sim\mathcal N(-s^2/2,s^2)$, completing the square gives
\begin{equation}
 \mathbb E[X^{-a}\mathbf1_{\{X\geq b\}}]
 =e^{a(a+1)s^2/2}
 \overline\Phi\!\left(\frac{\ln b+(a+1/2)s^2}{s}\right).
 \label{eq:truncated_moment}
\end{equation}
Using $F_P^{(0)}(p)=(p/A_0)^a$ on its support gives \eqref{eq:h2_reduced_pdf}--\eqref{eq:h2_reduced_cdf}. For $s=0$, the corresponding limit is $F_{\widetilde H_2}(u)=\min\{(u/B)^a,1\}$ for $u>0$; it is evaluated without dividing by $s$.

For $H=H_2Y$, the change $y=h/u$ has Jacobian $1/u$ for each $u>0$. Conditioning gives \eqref{eq:total_pdf}, and integrating the Weibull CDF gives \eqref{eq:total_cdf}. Alternatively, for $K>0$ the change $t=h/(\Lambda u)$ in the positive-density integral maps $u:0\to\infty$ to $t:\infty\to0$, with $|du|=h\,dt/(\Lambda t^2)$. Therefore
\begin{equation}
 f_H^{+}(h)=\frac K\Lambda\int_0^\infty
 t^{K-2}e^{-t^K}f_{H_2}^{+}\!\left(\frac h{\Lambda t}\right)dt.
 \label{eq:general_k_transform}
\end{equation}
The exponent remains $t^K$ for arbitrary $K$; $e^{-t^2}$ occurs only upon explicitly setting $K=2$. No $K=2$ expression is used to evaluate the general-$K$ law. Nonnegative integrands justify changing integration order, and the total probability is $p_0+(1-p_0)=1$.

\section{BER Transformation with a Possible Zero-Gain Atom}\label{app:ber}
For any nonnegative gain, including a mass at zero,
\begin{equation}
 Q(\sqrt{\gamma_0}H)=\frac1{\sqrt{2\pi}}
 \int_{\sqrt{\gamma_0}H}^{\infty}e^{-t^2/2}\,dt.
 \label{eq:q_integral}
\end{equation}
Interchanging this nonnegative integral and expectation yields
\begin{equation}
 P_b=\frac1{\sqrt{2\pi}}\int_0^\infty
 e^{-t^2/2}F_H(t/\sqrt{\gamma_0})\,dt.
 \label{eq:ber_tonelli}
\end{equation}
Putting $t=\sqrt2x$ proves \eqref{eq:ber_change}. This derivation avoids incorrectly dropping the endpoint mass in an integration-by-parts calculation. At zero SNR the limit is $1/2$; at unbounded reference SNR only the mass at $H=0$ contributes, giving \eqref{eq:ber_limit}. The limiting expressions provide consistency checks for the numerical evaluations.
\end{revision}

\section*{Acknowledgment}
The authors would like to thank the anonymous reviewers for their valuable comments.

\section*{Biography}
\begin{IEEEbiography}[{\includegraphics[width=1in,height=1.25in,clip,keepaspectratio]{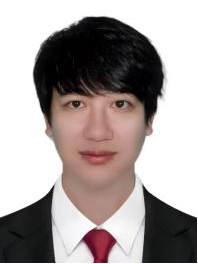}}]{Xing Du}
received the M.S. degree from Beijing Institute of Technology, China, in 2019, and the Ph.D. degree from the Aerospace Information Research Institute, Chinese Academy of Sciences, Beijing, China, in 2024. His research interests include free-space optical communication, channel modeling, and underwater laser communication technology. He was with the Computer, Electrical and Mathematical Sciences and Engineering Division, King Abdullah University of Science and Technology (KAUST). He is currently with the School of Information Science and Engineering, Southeast University.
\end{IEEEbiography}

\begin{IEEEbiography}[{\includegraphics[width=1in,height=1.25in,clip,keepaspectratio]{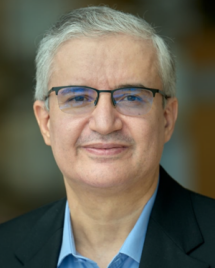}}]{Mohamed-Slim Alouini}
(Fellow, IEEE) was born in Tunis, Tunisia. He received the Ph.D. degree from California Institute of Technology (Caltech) in 1998. In 2009, he became a Founding Faculty Member with the King Abdullah University of Science and Technology (KAUST), where he is currently \rev{an} Al-Khawarizmi Distinguished Professor of electrical and computer engineering and the holder of the UNESCO Chair on Education to Connect the Unconnected. \rev{He was a faculty member} with the University of Minnesota and later with Texas A\&M University, Qatar. He is currently particularly focusing on addressing the technical challenges associated with information and communication technologies (ICT) in underserved regions and is committed to bridging the digital divide by tackling issues related to the uneven distribution, access to, and utilization of ICT in rural, low-income, disaster-prone, and hard-to-reach areas. His research interests encompass a wide array of research topics in wireless and satellite communications. He is a fellow of OPTICA.
\end{IEEEbiography}

\vfill
\end{document}